# Plasmonic Films Can Easily Be Better: Rules and Recipes

*Kevin M. McPeak, Sriharsha V. Jayanti, Stephan J. P. Kress,*

*Stefan Meyer, Stelio Iotti, Aurelio Rossinelli, and David J. Norris*[*]

Optical Materials Engineering Laboratory, ETH Zurich, 8092 Zurich, Switzerland.

ABSTRACT.  High-quality materials are critical for advances in plasmonics, especially as researchers now investigate quantum effects at the limit of single surface plasmons or exploit ultraviolet- or CMOS-compatible metals such as aluminum or copper. Unfortunately, due to inexperience with deposition methods, many plasmonics researchers deposit metals under the wrong conditions, severely limiting performance unnecessarily. This is then compounded as others follow their published procedures. In this perspective, we describe simple rules collected from the surface-science literature that allow high-quality plasmonic films of aluminum, copper, gold, and silver to be easily deposited with commonly available equipment (a thermal evaporator). Recipes are also provided so that films with optimal optical properties can be routinely obtained.

Plasmonics requires simple methods to deposit metallic films with optimized optical performance and precise microstructure. Such films allow low-loss plasmonic devices with well-placed nanoscale features to be created. However, for a given metal, the dielectric function (or relative permittivity) and film morphology depend greatly on the deposition conditions. Many plasmonics researchers struggle with this problem, particularly if film growth is not within their area of expertise. Fortunately, surface scientists have reported extensively over the past 60 years on the relationships between how a metallic film is deposited and its properties.[1-4] Here, we aim to condense this knowledge into a useful form for the plasmonics community. We discuss the key issues and then provide recipes that can be used to improve the properties of plasmonic films without any additional effort.

Experimentally, the goal is to deposit dense metallic films with high purity, low surface roughness, and large grain sizes (*i.e.* grain diameters of approximately the film thickness). In such films, losses are reduced and precise structures can be formed with focused-ion-beam (FIB) lithography or other techniques. To grow these films, here we assume the reader has access to a commonly available deposition apparatus, a thermal evaporator. Further, we assume that the evaporator is limited to high-vacuum conditions (with pressures in the $10^{-6}$ to $10^{-8}$ Torr range) instead of ultra-high vacuum. For simplicity, we consider standard silicon wafers (including their native oxide) as the substrates with the deposition performed at room temperature. This temperature allows simple evaporators without substrate heaters to be utilized. It also avoids undesirable dewetting of the metal from the substrate, which can occur at elevated temperatures, as discussed below. Under these conditions, we then exploit prior literature knowledge and show how to deposit four common plasmonic metals (Al, Cu, Au, and Ag). Silver and gold are well-known plasmonic metals. Aluminum and copper are becoming increasingly important for ultraviolet (UV) and CMOS-compatible applications.[5-8]

**RULES**

In general, many experimental factors affect the optical performance and microstructure of pure polycrystalline metallic films. This includes the deposition rate, the base pressure in the deposition chamber, the substrate temperature, and the overall film thickness. How sensitive the final behavior of the film is to these experimental parameters also depends on the intrinsic properties both of the metal (*e.g.*, its reactivity, bulk- and surface-diffusion coefficients, and surface energy) and of the substrate (*e.g.*, its roughness and surface energy). The question we wish to address is which of these factors is the most important for optimizing the deposition of plasmonic metallic films. In other words, what rules should we follow?

Previous work by surface scientists has shown[1-4] that the homologous temperature of the deposition ($T_h$), which is defined as the ratio of the substrate temperature, $T_{sub}$, to the melting temperature of the metal, $T_{mp}$, is useful for describing the growth mechanism and resulting microstructure in a variety of thin films. Specifically, $T_h$ provides an effective scale for describing fundamental thermally activated phenomena in film formation (*e.g.* nucleation, crystal growth, and grain growth). However, one can imagine that additional factors, such as the reactivity of the metal (*e.g.* how easily it reacts with residual gases in the chamber)[9,10] as well as the metal-substrate contact angle[11,12] (*e.g.* wetting) will also strongly affect the purity, density, and hence quality of the deposited film.

We reduce the complexity of all of these factors by representing each metal in terms of just two key parameters. This oversimplifies many of the details of the deposition, but is sufficient for our purposes. Specifically, Figure 1 plots the homologous temperature of the deposition versus the standard electrode potential for Al, Cu, Au, and Ag. Metals with higher standard electrode potentials are more reactive, as indicated qualitatively by the arrow at the top of the graph.

What can we learn from such a plot? First, all of the metals studied here are deposited with $T_h$ ~ 0.3 (assuming $T_{sub}$ is at room temperature). Surface science teaches that films deposited with $0.15 < T_h < 0.3$ contain metastable phases with surface-diffusion-driven grain growth proceeding for the mobile grain boundaries.[2] Growth models refer to this range of $T_h$ as a transition zone since more than one but not all grain boundaries are mobile.[2,3] This results in a bimodal distribution in grain size. Indeed, this effect can be observed in the scanning electron micrographs of our deposited films (see Figure 2). Considering that all of our metals have similar $T_h$ values, it is not surprising that their grain sizes are similar, in the 500 nm to 1 μm range. Perhaps more importantly for our present goals, their similar $T_h$ values indicate that it is reasonable to pick a fixed deposition temperature for all four metals (*e.g.* room temperature) and not consider this parameter further for their optimization. This is the first rule learned from Figure 1.

Second, Figure 1 shows that our metals have a range of reactivities. This parameter is important to consider because it describes how readily the metal atoms will interact with residual gas molecules in the chamber (*i.e.* gas molecules that have not been removed by the vacuum pump). The dominant residual gas in the high-vacuum regime is water vapor.[13] Oxygen typically has a partial pressure an order of magnitude lower, but is more reactive.[14] Even for metals with low reactivity, these residual gases can be problematic. They adsorb on the freshly deposited film and pin grain boundaries, which reduces the average grain size in the film, thus creating more electron-scattering centers.[10,15] This can be particularly harmful to the optical properties of metallic films in the UV and visible regimes due to their large number of free electrons.[10,16] Thus, the second rule from Figure 1 is that we should always deposit under the best vacuum conditions possible to reduce the effects of residual gases. (Acceptable pressure ranges for each metal are listed in the Recipes section below.)

However, for reactive metals, our best vacuum may not be sufficient. In this case, the metal atoms can still react with trace water or oxygen molecules in the chamber before depositing on the substrate. The film will then be contaminated with metal oxide. To avoid this, we should deposit reactive metals at fast rates to reduce the interaction time. Furthermore, ultrafast deposition of reactive metals offers the added benefit that the concentration of residual gases can actually be reduced (or gettered) by reactions in the chamber before deposition begins (*i.e.* before the substrate shutter is opened). Thus, the third rule from Figure 1 is that we should increase the deposition rate for reactive metals.

Even for Ag, which is not particularly reactive but is susceptible to grain-boundary pinning by residual gases (mentioned above), fast deposition rates can improve film quality. For example, Figure 3 shows atomic force micrographs for three Ag films deposited under our best vacuum conditions ($3\times10^{-8}$ Torr), but at different rates. Even at this low pressure, the deposition rate of 25 Å/sec results in a significant increase in grain size compared to the more commonly used 1 Å/sec. Figure S1 in the Supporting Information confirms a gradual reduction in optical losses as the rate is increased, consistent with reduced electron scattering in films with larger grain size.

Should one conclude from the above discussion that fast deposition rates are always better? Fast rates can also potentially have negative consequences. In addition to wasting material during ramp-up of the evaporator (which for expensive metals is a relevant issue), fast rates also reduce the time that freshly added metal atoms (adatoms) can diffuse on the substrate before being bombarded by additional atoms. This has the tendency to produce smaller grains, especially at low substrate temperatures where surface diffusion is minimal.[2,17] However, in the high-vacuum regime, where the amount of residual gases can be considerable, the ability of these contaminants to pin grain boundaries and form metal-oxide inclusions can be a larger concern, depending on the metal.

Based on all of these considerations, we have added a graded background color in Figure 1 to provide a qualitative guide for the relative deposition rates necessary to obtain a particular metal with good optical properties and low roughness under the assumed experimental conditions. Highly reactive metals, such as Al, should be deposited at 100 Å/sec or faster.[16] Unreactive metals such as Au are far less susceptible to residual gases and therefore can be deposited at lower rates (*e.g.* 1 Å/sec) without deleterious effects to their optical properties. We note, however, that for a given metal, the required deposition rate can change depending on the vacuum conditions (*i.e.* the amount of residual gases). We give examples of this below.

Following this discussion, it is interesting to now return and ask whether substrate heating would improve the films further. In other words, above we followed our first rule and fixed the substrate temperature for all of our metals at room temperature for experimental simplicity. Would we gain in performance by increasing the substrate temperature (and hence $T_h$)? Higher $T_h$ can certainly increase grain size by providing more energy for surface diffusion. Larger grain sizes could potentially be a benefit for plasmonic applications. Unfortunately, elevated temperatures can also result in dewetting of the metastable film.[12] Dewetting is a process driven by minimization of surface energy and can occur at temperatures well below the melting point of the material.[12] Noble metals deposited on oxide-coated Si are particularly susceptible to this effect since weak adhesion at the metal-substrate interface lowers the barrier for dewetting.[18] Ag films in particular have shown dewetting phenomena such as pinhole formation and aggregation at temperatures as low as 100 °C in our laboratory and by others.[19] This can lead to significant increases in porosity and roughness in the film. Thus, room-temperature deposition can actually provide a better outcome in such cases.

If room temperature is beneficial, one might also worry that under fast growth rates the substrate temperature will be increased through heat-transfer processes during the deposition. This could lead to the detrimental dewetting phenomena just discussed. In fact, fast deposition rates actually do the opposite; they help minimize substrate heating. While seemingly counterintuitive, radiant heating from long deposition times at slow rates in a thermal evaporator typically results in a larger increase in the substrate temperature than from short deposition times at fast rates.[20] Furthermore, the heat of condensation, which is released when a solid metal film is deposited on the substrate, does not begin to contribute significantly to substrate heating until the evaporation rate is in excess of 500 Å/sec, even for highly reactive metals such as Al.[20] Such rates are more than three times faster than the fastest rates used in the recipes given below. Therefore, fast deposition rates not only help mitigate contamination but also help avoid undesirable substrate heating, which can increase film roughness through dewetting.

We now summarize the key rules for depositing metallic films for plasmonics via a simple thermal evaporator. Residual gases should be avoided with the lowest pressure one can attain. One should not expect good optical films from a poor vacuum (pressures above $10^{-5}$ Torr). In the high-vacuum regime, one can deposit metals with optimized optical properties. However, the more reactive the metal, the faster the deposition rate should be to combat the detrimental effects of residual gases. For less reactive metals, we must select a deposition rate that balances grain growth and grain-boundary pinning caused by the residual gases. Finally, although heated substrates could potentially increase grain size during growth, this benefit is frequently outweighed by undesirable dewetting effects in the film, especially for Ag. Thus, the example films described below are obtained from room-temperature deposition, which is also amenable to a simple apparatus.

**OPTICAL PERFORMANCE**

Above we presented rules to give the reader intuition about the deposition process. If these rules are followed in practice, how do they impact optical performance? In the next section we list recipes that are based on these rules (*i.e.* they follow the prior surface-science literature). We used them to deposit 300-nm thick films of Al, Cu, Au, and Ag on native-oxide-coated Si wafers. To have the smoothest surfaces possible for accurately extracting optical properties, all films were then template-stripped from their substrates.[21,22] After deposition, an adhesive and counter-substrate were added to the top "as-deposited" surfaces of the films and they were peeled off the wafer, exposing the metal interface that initially formed at the native oxide. The measured root-mean-squared (RMS) surface roughness of the as-deposited surfaces was typically a few nanometers. The values reported below for the template-stripped surfaces are significantly lower. (Figure 2 demonstrates Al, Cu, Au, and Ag films that were deposited according to the recipes and then template stripped from wafers that were first structured via FIB lithography.) Template-stripping is also beneficial for providing pristine films (*i.e.* avoiding contamination) as the metal interface can be protected until the last moment. We template stripped our films immediately prior to optical measurements, which were then performed under flowing $N_2$ gas. Atomic force microscopy (AFM) images were collected within the hour.

Optical properties were obtained with a variable-angle spectroscopic ellipsometer (V-VASE or VUV-VASE, J. A. Woollam Co.). Because of the smooth template-stripped surfaces, we could exploit a simple two-layer vacuum-metal model to extract the dielectric function of each metal (except for Al, where an oxide layer was also included). If we instead measured the as-deposited surfaces, which are rougher, the fit of the ellipsometry data would be less reliable. Further, our goal is not to address the impact of residual roughness on the effective dielectric functions obtained by ellipsometry, but to summarize how deposition conditions affect the dielectric function of each metal. Finally, we note that we explored a range of pressures and deposition rates, and observed significant changes in the optical performance, even on the smooth template-stripped surfaces (see details below). Our results are in agreement with the prior surface-science literature and the rules summarized above.

Experimental data for the recipe films that show the real and imaginary components of the dielectric functions (solid lines) are plotted in Figure 4a,b, respectively. (Digital files containing the actual data points are provided in the Supporting Information.) Figure 4a,b also shows literature values (dashed lines) from Palik[23] (Cu, Au, and Ag) and Rakić[24] (Al) for comparison. Palik and Rakić were chosen as standards due to their widespread use in the plasmonic community. (We plot our results against another common standard, Johnson and Christy,[25] in Figure S2 in the Supporting Information.) The real components ($\varepsilon'$) of the dielectric functions for the recipe films of Cu, Au, and Ag agree well with Palik, while their imaginary components ($\varepsilon''$), which are related to losses, are significantly better. For Al, our values are very similar to those from Rakić. These results show that by following the rules discussed above, films with optimal optical performance can be obtained even with a simple apparatus. Such films exhibit properties significantly better than those obtained under deposition conditions commonly used in the plasmonics community (*i.e.* at slow rates). Indeed, they have properties better than the standard literature values.

Is it surprising that the recipe films have better dielectric functions than in the standard references? Within this context, it is important to point out that the standard values[23-25] were obtained from films deposited according to the rules summarized above, which is another indication of their validity. However, these films were typically a factor of ten thinner than the films presented here. Under similar deposition conditions, thinner films will have smaller grains. Thus, our thicker films can exhibit smaller losses due to reduced grain-boundary scattering. The standard references may have also exploited thinner films because they can provide smoother as-deposited surfaces. Unfortunately, the surface roughness values were not reported. If their films were rougher than our template-stripped surfaces, this could also affect the reported dielectric functions. In either case (due to smaller grain size or increased roughness), the data in Figure 4a,b indicate that the standard reference values should not be used to simulate perfect, ideal plasmonic structures. Better values can easily be obtained in practice.

The impact of these improvements is demonstrated in Figure 4c,d, where two figures-of-merit[26] are plotted. Quality factors ($Q$s) for our films (solid lines) are compared with literature values (dashed lines) for both localized surface plasmon resonances (LSPR) and surface plasmon polaritons (SPPs). Table 1 shows the percentage improvement in the $Q$s at selected wavelengths for the different metals.

Clearly, significant boosts in performance are possible. The message of this work is that such an improvement can easily be achieved, in fact without any additional experimental effort. The deposition simply needs to be performed under the appropriate conditions.

We now discuss the optical performance of each of the metals separately.

*Silver* Ag is perhaps the most widely used plasmonic material due to its low losses in the visible regime. Because it is a noble metal, it is often considered to be a material that is easy to deposit via thermal evaporation. However, we caution that the optical properties of Ag films can be significantly reduced by water vapor and oxygen in the vacuum chamber.[10] In fact, in many respects, Ag is the most unforgiving of the four metals to deposit. As already mentioned, Ag is susceptible to dewetting and grain-boundary pinning. For example, compared to Cu, which is at a similar position in Figure 1, the adhesion energy of Ag on silica is nearly 50% lower,[27] leading to a tendency of the metal to dewet or "ball up," even at room temperature. Another complication is the crystalline orientation of the grains in the film, a factor that we ignored for simplicity in our rules above. For Ag, template-stripped films with the desired (111)-oriented grains are obtained only for pressures around $1\times10^{-7}$ Torr or better.[28] Above these pressures, random grains with irregular grain boundaries are formed. [For comparison, the (111) orientation dominates in Au all the way up to $1\times10^{-5}$ Torr.] Therefore, due to all of these effects, Ag films with pinholes and increased grain structure can easily form under poor vacuum conditions or at rates that are too slow. Figure S3 in the supporting information highlights these effects showing two AFM images of template-stripped Ag deposited at a slow rate of 0.2 Å/sec with a base pressure of $2\times10^{-6}$ Torr (Figure S3a) and $3\times10^{-8}$ Torr (Figure S3b). The film deposited at $2\times10^{-6}$ Torr has a roughness of 1.23 nm RMS whereas the film deposited at $3\times10^{-8}$ Torr has a roughness of 0.32 nm RMS.

Under our best vacuum conditions ($3\times10^{-8}$ Torr) we deposited Ag at rate of 50 Å/sec to obtain films that are significantly better than Palik,[23] showing at least a 160% improvement in the LSPR figure-of-merit and at least a 200% improvement in the SPP figure-of-merit, over a large spectral range (see Table 1). The surface roughness was 0.37 nm RMS (see Figure S4a). When our vacuum was an order of magnitude worse ($3\times10^{-7}$ Torr), we could still obtain films of similar or better surface roughness (0.30 nm RMS) and optical quality (see Figure S5), but only by increasing our deposition rate to 150 Å/sec, consistent with the rules above. At even higher pressures, the quality of the Ag films deteriorates significantly. Thus, for Ag, a good vacuum is required.

*Copper* Cu is a low-cost CMOS compatible metal that is slightly more reactive than Ag (Figure 1) and thus, according to the rules, slightly more susceptible to residual-gas contaminants during deposition. Ideally, we should increase the deposition rate compared to Ag. However, the fastest attainable deposition in our evaporator was 35 Å/sec (due to Cu creep, see Recipes section). We used this rate at a base pressure of $3 \times 10^{-8}$ Torr. The resulting materials exhibit a 30-100% improvement over Palik for both the LSPR and SPP figures-of-merit (see Table 1). Perhaps more remarkable is that such Cu films outperform our best Au films for most of the near-infrared (near-IR) and specifically at 1550 nm. The predicted SPP propagation length at 1550 nm on our Cu film is 820 μm. This is significant given recent interest in Cu for CMOS-compatible plasmonic interconnects.[5] The surface roughness of the Cu films was 0.25 nm RMS (see Figure S4b).

We note that, while Figure 1 indicates that Cu is slightly more reactive than Ag, in practice Cu requires much less stringent deposition conditions in comparison. This can be due to a combination of the effects already discussed above for Ag. Cu has a weaker tendency to dewet[27] leading to a less-complicated grain structure. It has also been reported that (111)-oriented grains do not dominate in deposited Cu until pressures below $1 \times 10^{-8}$ Torr.[28] Thus, under our conditions, the grains are randomly oriented. For Cu, the good adhesion and random grains lead to smooth films with good optical performance. Indeed, when we deposited Cu at a base pressure of $3 \times 10^{-7}$ Torr at 25 Å/sec we obtained essentially the same optical properties and roughness as at $3 \times 10^{-8}$ Torr (see Figure S6).

*Gold* Au is another widely used plasmonic metal, particularly in bio-related devices at red wavelengths.[29-31] The inertness of Au means that residual gases have far less impact on its optical properties. That said, the Au-recipe films outperform Palik by 30 to 95% in the red and near-IR regime for both LSPR and SPP figures-of-merit (see Table 1). These films were deposited at a base pressure of $3 \times 10^{-8}$ Torr and 10 Å/sec. The surface roughness was 0.3 nm RMS (see Figure S4c). Films that were deposited at $2 \times 10^{-6}$ Torr and 0.5 Å/sec had optical properties marginally worse and marginally better in the visible and IR, respectively (see Figure S7). The surface roughness increased to 0.4 nm RMS.

*Aluminum* Al is considered the best plasmonic metal for the UV[7,8] and has recently been of interest for metal-enhanced fluorescence,[32] deep-UV Raman scattering,[33,34] non-linear plasmonics,[35] high-energy plasmonics,[36] and CMOS-compatible color filters.[6] Unfortunately, Al is also extremely reactive (Figure 1) and therefore highly sensitive to residual gases in the deposition chamber. Following the work of Hass,[16] we found that extremely fast evaporation rates (~150 Å/sec) were necessary to closely match the LSPR and SPP figures-of-merit from Rakić in the UV. Ultra-fast deposition rates result in more compact films that are also less susceptible to oxidation over time.[16] We stress that the optical properties suffered significantly in both the UV and visible regimes at more commonly used deposition rates of 1 to 5 Å/sec, even if the chamber pressure was as low as $3 \times 10^{-8}$ Torr (see Figure 5). Indeed, under such commonly used deposition rates, our data show that the SPP propagation lengths in the UV were 80% less than predicted by the data from Rakić. Films deposited at slower rates were also not template strippable presumably due to strong adhesion between oxidized aluminum and the native oxide on the silicon wafer. Therefore, in the high-vacuum regime extremely fast deposition rates are critical to achieve high-quality Al films. Under these conditions, we also achieved a surface roughness of 0.58 nm RMS (see Figure S4d).

**RECIPES**

All films were deposited in a Kurt J. Lesker Nano36 thermal evaporator equipped with the standard 3.3 V, 375 A power supply and dual source/substrate shutters. The source-to-substrate-distance for all deposition runs was 30 cm. The chamber was pumped with a 685 L/sec turbo pump. A custom-built Meissner trap was also installed in the chamber to aid water-vapor removal and decrease pump-down times of the deposition chamber. It consisted of two copper plates (roughly 30 by 30 cm) cooled by a coiled copper tube that was filled with liquid $N_2$. The trap was required for our chamber to achieve our lowest base pressure of $3 \times 10^{-8}$ Torr.

All films were deposited on native-oxide-covered Si(100) wafers. These substrates were cleaned with 10 min of sonication in both acetone (Univar AG) and isopropyl alcohol (Thommen Furler AG), followed by 10 min of sonication at 45 °C in RCA cleaning solution, which contained 20 mL of 30% hydrogen peroxide (VWR Chemicals, AnalaR NORMAPUR), 4 mL of 30-32% aqueous ammonium hydroxide (Sigma Aldrich, ACS reagent), and 100 mL of H$_2$O (deionized by a MilliQ Advantage A10 System, 18.2 MΩcm at 25 °C).

*Silver* A 49-mm-long by 12-mm-wide by 0.4-mm-thick tungsten dimple boat (Umicore) was used with 1/8-inch by 1/8-inch 99.99% Ag pellets (Kurt J. Lesker) as the source material. The base pressure in the chamber was $3 \times 10^{-8}$ Torr. The deposition rate was 50 Å/sec. We also tested deposition at a base pressure of $3 \times 10^{-7}$ Torr. In this case, to maintain the quality of the film, a faster deposition rate of 150 Å/sec was required (see Figure S5).

*Copper* A 4-inch-long by 0.5-inch-wide by 0.015-inch-thick tungsten dimple boat (R. D. Mathis) was used with 1/8" x 1/8" 99.99% Cu pellets (Kurt J. Lesker) as the source material. The base pressure in the chamber was $3 \times 10^{-8}$ Torr. The deposition rate was 35 Å/sec. Similar results were obtained at $3 \times 10^{-7}$ Torr and 25 Å/sec (see Figure S6). Note that Cu will creep considerably on the tungsten boat during deposition. If too much power is applied, the Cu can reach the electrodes and result in shorting. This limited our fastest deposition rate to 35 Å/sec.

*Gold* A 49-mm-long by 12-mm-wide by 0.4-mm-thick tungsten dimple boat (Umicore) was used with 1/8-inch by 1/8-inch 99.999% Au pellets (ACI Alloys) as the source material. The base pressure in the chamber was $3 \times 10^{-8}$ Torr. The deposition rate was 10 Å/sec. Films were also deposited at $2 \times 10^{-6}$ Torr and 0.5 Å/sec with minor changes in the properties (see Figure S7).

*Aluminum* A 4-inch-long by 0.5-inch-wide by 0.01-inch-thick tungsten dimple boat (R. D. Mathis) was used with 1/8-inch by 1/8-inch 99.999% Al pellets (Kurt J. Lesker) as the source material. Thinner boats are not advised due to alloying between aluminum and tungsten at elevated temperatures. Base pressures in the chamber of $1 \times 10^{-6}$ Torr and below were used successfully for Al deposition. Once the base pressure was reached, the boat was heated slowly (~5 to 10 min ramp) until the Al pellets melted. The power was gradually increased until a slow rate (~0.1 Å/sec) of metal deposition was detected on the quartz crystal microbalance. The power was then increased quickly by at least 50% (without exceeding the current limit of the power supply). The Al spread quickly across the tungsten boat. The rate on the quartz crystal microbalance was monitored and the substrate shutter was opened when a rate of ~150 Å/sec or greater was reached. Rates even higher than 150 Å/sec did not show improved optical properties but did result in increased roughness (*e.g.* 0.58 nm RMS for 150 Å/sec and 0.8 nm RMS for 400 Å/sec).

**CONCLUSION**

We have presented a series of rules and recipes to aid researchers in depositing plasmonic metallic films with optimized structural and optical properties. We have restricted our discussion to room-temperature deposition with a standard thermal evaporator. In this case, the primary experimental parameters to consider are the base pressure in the vacuum chamber and the deposition rate. One should always use the lowest pressure possible; good optical films cannot be expected from poor vacuum conditions (pressures above $10^{-5}$ Torr). Even under good vacuum conditions ($10^{-8}$ Torr), reactive metals such as aluminum require fast deposition rates to avoid metal-oxide contamination. Less reactive metals require a deposition rate that balances grain growth and grain-boundary pinning. Of course, due to our focus on thermal evaporation, we have not addressed more sophisticated deposition strategies that have recently been explored in plasmonics to improve material quality. These have included the growth of single-crystalline flakes and films.[37-42] By avoiding grain structure, such films can allow more precise patterning of plasmonic structures, which is clearly beneficial. However, these techniques, while not overly difficult to implement in the laboratory, do require additional experimental capabilities beyond most optics labs, such as high-temperature sputtering or low-temperature molecular beam epitaxy. In our experience, the improvement in optical properties over what one can obtain with a well-deposited polycrystalline film is also marginal. Thus, for many experiments in plasmonics, the very simple approach presented here is sufficient.

**Corresponding Author**

*Email: dnorris@ethz.ch

**Notes**

The authors declare no competing financial interest.

**Supporting Information**. Further experimental details and additional data related to film characterization via ellipsometry and atomic force microscopy. This material is available free of charge via the Internet at http://pubs.acs.org.

ACKNOWLEDGMENTS

We gratefully acknowledge assistance from Marilyne Sousa with the ellipsometry measurements. The research leading to these results received funding from the European Research Council under the European Union's Seventh Framework Programme (FP/2007-2013) / ERC Grant Agreement Nr. 339905 (QuaDoPS Advanced Grant) and from the Swiss National Science Foundation under award no. 200021-140617.

**TABLES**

**Table 1.** Percentage increase in the figures-of-merit[26] for Al, Cu, Au, and Ag at ultraviolet (280 nm), visible (650 nm), near-infrared (1000 nm), and telecommunication (1550 nm) wavelengths. The increase is based on comparison to Palik[23] (Cu, Au, and Ag) and Rakić[24] (Al). The quality factors for localized surface plasmon resonances ($Q_{LSPR}$) and surface plasmon polaritons ($Q_{SPP}$) are shown along with the calculated surface plasmon propagation lengths ($L_{SPP}$) based on the measured dielectric functions of the recipe films. Specifically, the films were deposited under the conditions detailed in Figure 2. Very similar films could be obtained over a range of conditions, as described in the text.

| Wavelength Regime | Metal | Increase over Standard References[23,24] | | $L_{SPP}$ (µm) |
| --- | --- | --- | --- | --- |
| | | $Q_{LSPR}$ (%) | $Q_{SPP}$ (%) | |
| Ultraviolet (280 nm) | Al | 11 | -12 | 2.5 |
| Visible (650 nm) | Ag | 200 | 250 | 84 |
| | Cu | 120 | 130 | 24 |
| | Au | 32 | 38 | 20 |
| Near-Infrared (1000 nm) | Ag | 160 | 200 | 340 |
| | Cu | 100 | 93 | 190 |
| | Au | 51 | 61 | 190 |
| Telecom (1550 nm) | Ag | 270 | 480 | 1200 |
| | Cu | 140 | 120 | 820 |
| | Au | 81 | 95 | 730 |

**FIGURES**

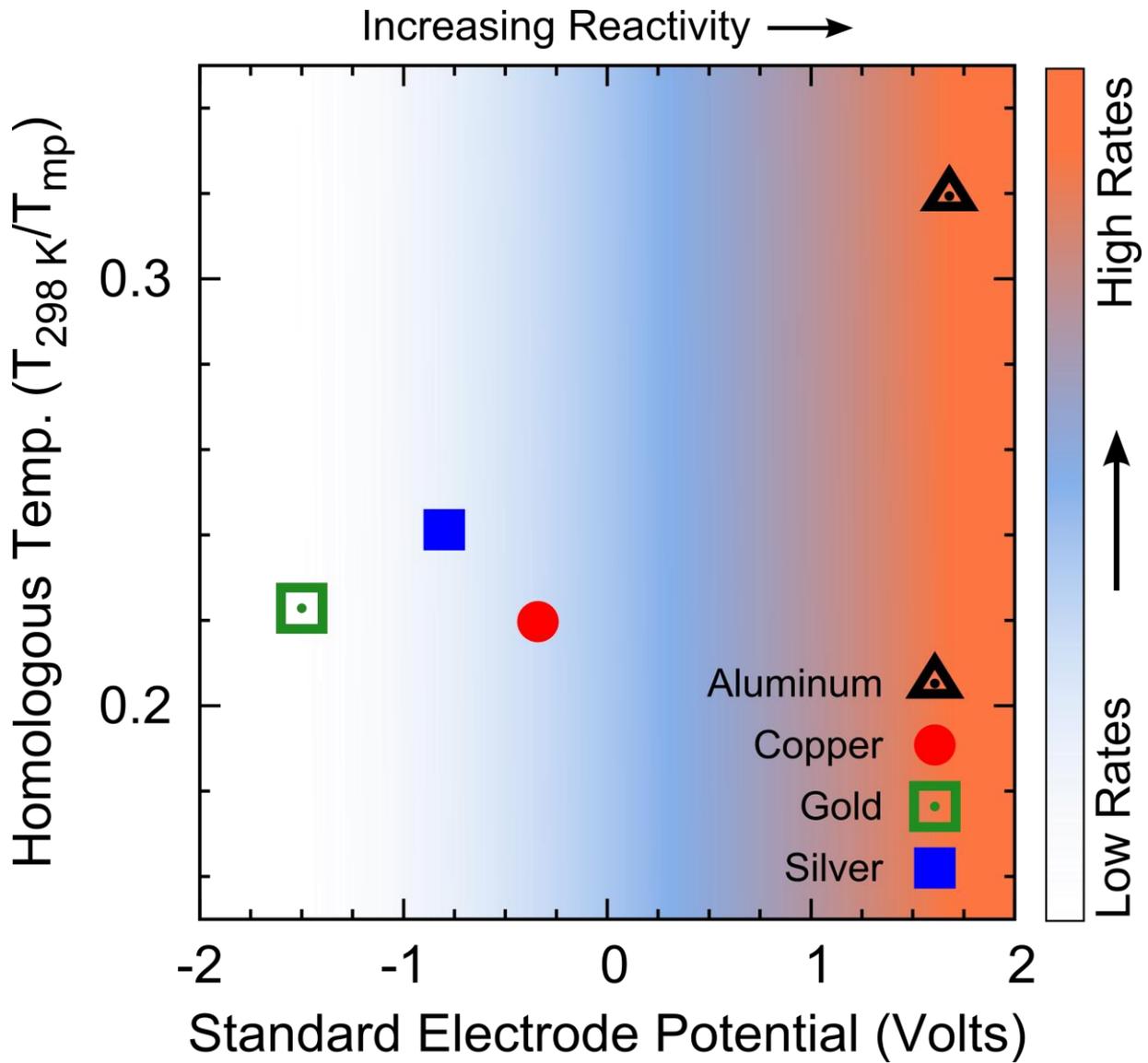

**Figure 1.** Homologous temperature ($T_{sub}/T_{mp}$ in Kelvin) versus standard electrode potential in volts for Al (black), Cu (red), Au (green), and Ag (blue). The reactivity of the metal increases to the right. The background color represents the qualitative trend in deposition rates necessary to achieve high-quality metallic films in the high-vacuum regime, as discussed in the text.

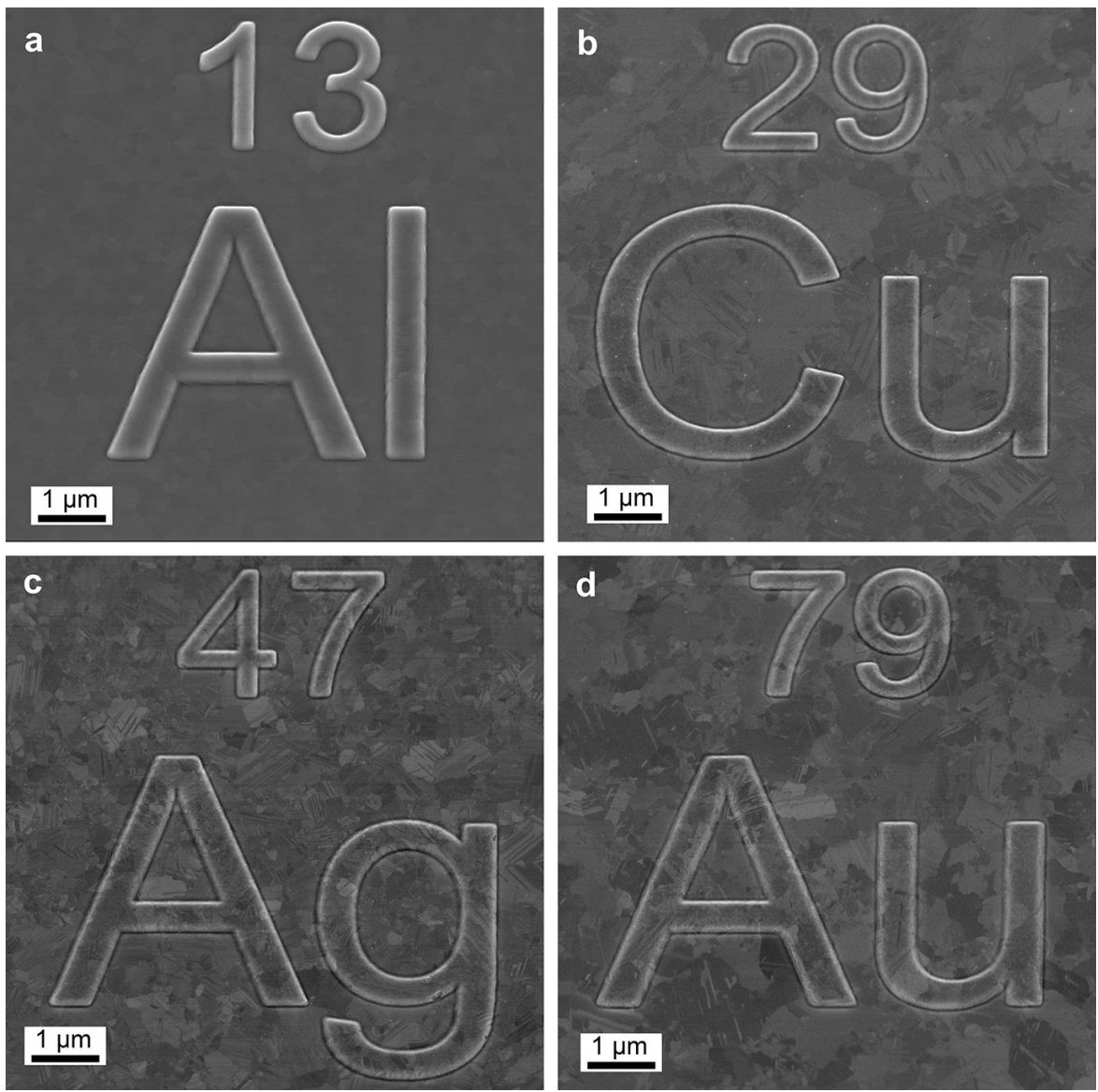

**Figure 2.** Scanning electron micrographs of structured films of (a) Al, (b) Cu, (c) Ag, and (d) Au, which were template-stripped from Si templates pre-patterned by focused-ion-beam lithography. The depositions were performed at room temperature at a base pressure of $3 \times 10^{-8}$ Torr and rates of 150, 35, 50, and 10 Å/sec, respectively.

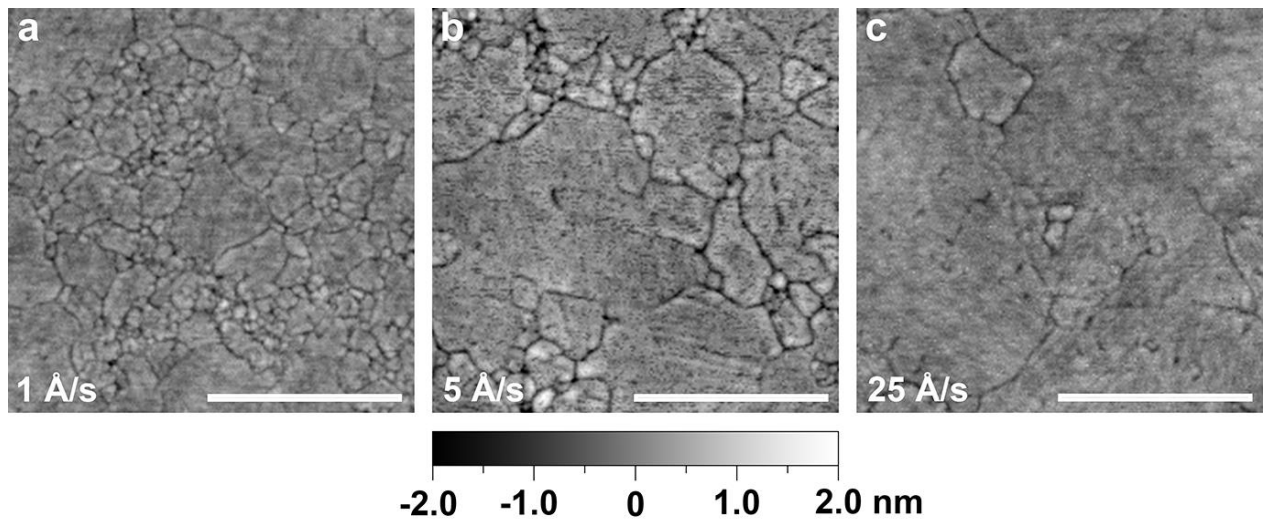

**Figure 3.** Atomic force micrographs for template-stripped silver films deposited at a base pressure of $3 \times 10^{-8}$ Torr but different deposition rates: (a) 1, (b) 5, and (c) 25 Å/sec. The roughness values for the films over a 2.5 μm x 2.5 μm area are: (a) 0.32, (b) 0.45, and (c) 0.32 nm RMS. While all the films have similar roughness, their grain size increases with faster deposition rates. This results in a reduction in the optical losses, as shown in the Supporting Information (Figure S1). The scale bar corresponds to 500 nm.

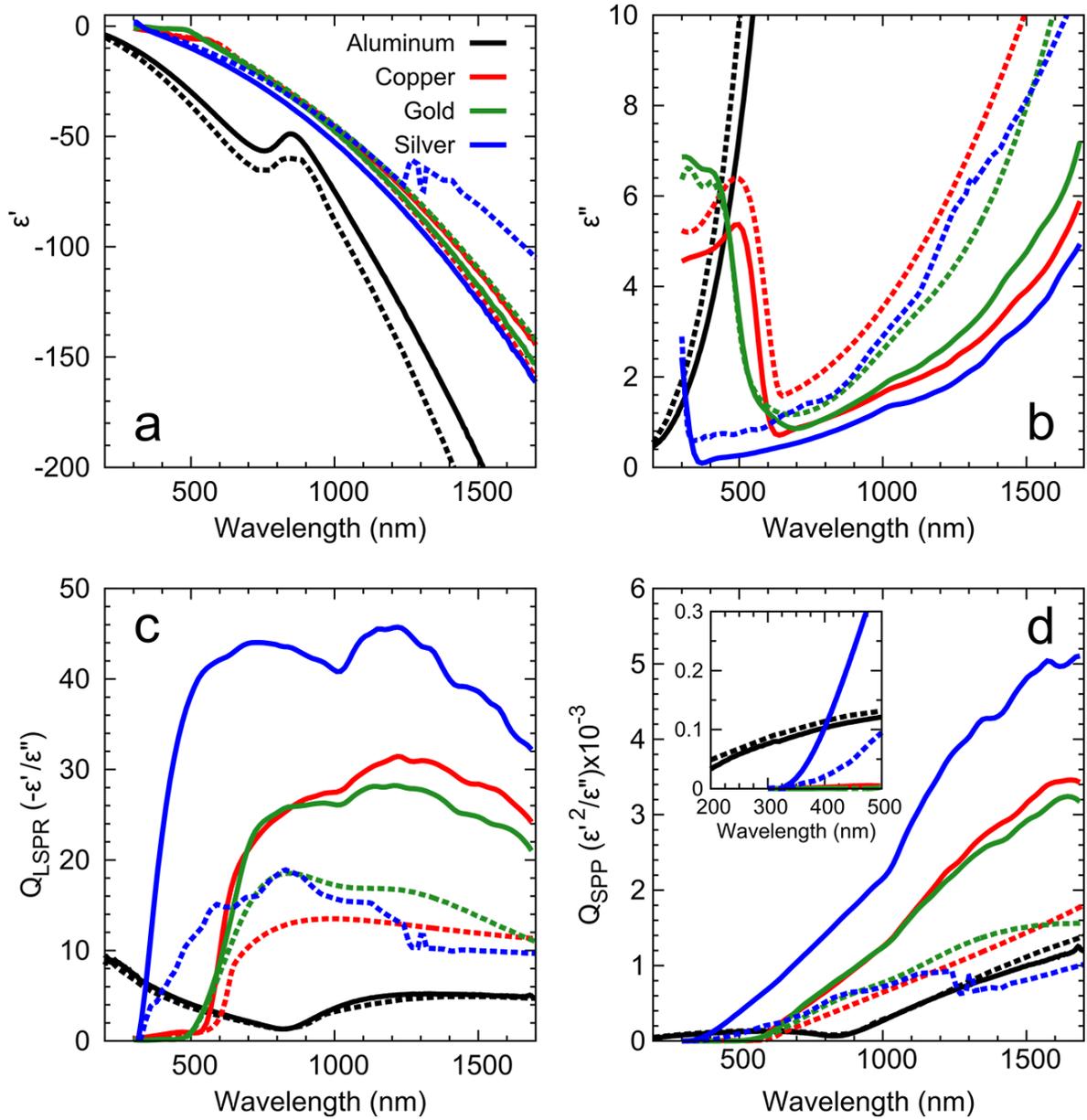

**Figure 4.** Optical properties and figures-of-merit for our metallic films of Al (black), Cu (red), Au (green), and Ag (blue). The films were deposited as described in Figure 2. Solid lines are for measured values for the template-stripped recipe films and dashed lines are from Palik[23] (Cu, Au, and Ag) and Rakić[24] (Al). (a) and (b) show the real and imaginary part of the dielectric function, respectively. (c) and (d) plot calculated quality factors for the localized surface plasmon resonance in spherical structures ($Q_{LSPR}$) and surface plasmon polaritons ($Q_{SPP}$), respectively. The data for the films were smoothed with a five-point moving average.

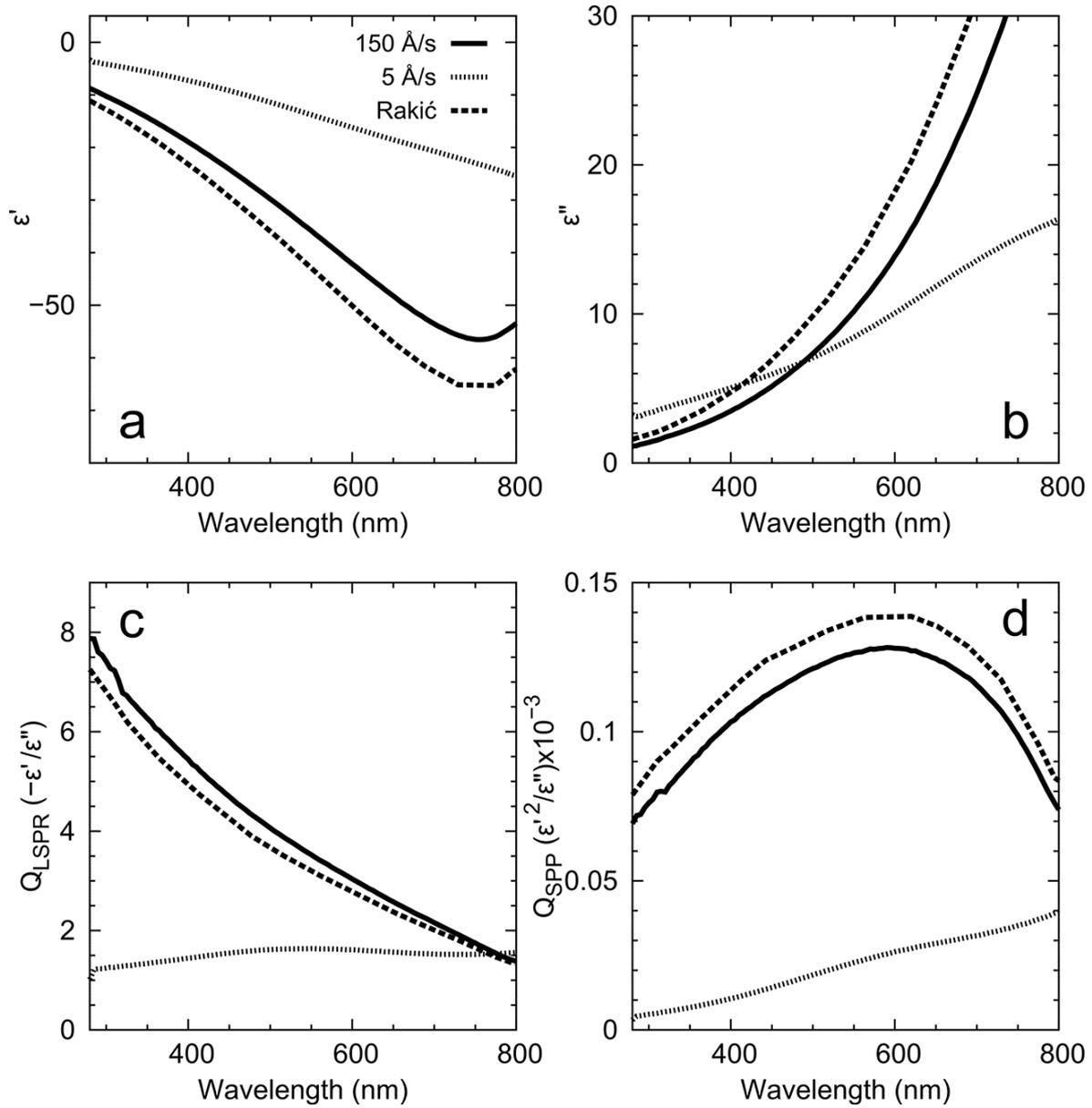

**Figure 5.** Optical properties and figures-of-merit for Al films deposited at room temperature at a base pressure of 3x10$^{-8}$ Torr. Solid and dotted lines compare properties for a template-stripped film deposited at a fast rate (150 Å/sec) and the as-deposited surface from a film grown at a rate more typically found in the plasmonics literature (5 Å/sec), respectively. Despite the additional roughness in the latter case, the measured dielectric functions are dominated by the presence of aluminum oxide in the film. Note how the fast rates are consistent with Rakić[24] (dashed lines). (a) and (b) show the real and imaginary part of the dielectric function, respectively. (c) and (d) plot calculated quality factors for the localized surface plasmon resonance in spherical structures ($Q_{LSPR}$) and surface plasmon polaritons ($Q_{SPP}$), respectively.